\documentclass[pre,aps,showpacs,twocolumn,floatfix,superscriptaddress]{revtex4}

\usepackage{graphicx}
\usepackage{bm} 
\usepackage{epsfig}
\usepackage{xcolor}

\newcommand{\be}{\begin{equation}}
\newcommand{\ee}{\end{equation}}
\newcommand{\bea}{\begin{eqnarray}}
\newcommand{\eea}{\end{eqnarray}}
\newcommand{\ba}{\begin{array}}
\newcommand{\ea}{\end{array}}

\begin{document}
\title{End-pulled polymer translocation through a many-body flexible pore}

\author{A. Fiasconaro}
\email{afiascon@unizar.es}

\affiliation{Dpto. de F\'{\i}sica de la Materia Condensada,
Universidad de Zaragoza. 50009 Zaragoza, Spain}

\affiliation{Instituto de Biocomputaci\'on y F\'{\i}sica de Sistemas
Complejos, Universidad de Zaragoza. 50018 Zaragoza, Spain}

\author{G. Diez-Se\~norans}

\affiliation{Grupo de Dise\'no Electrónico, Dpto. de Ingenier\'ia Electr\'onica y Comunicaciones,
Universidad de Zaragoza. 50009 Zaragoza, Spain}

\author{F. Falo}

\affiliation{Dpto. de F\'{\i}sica de la Materia Condensada,
Universidad de Zaragoza. 50009 Zaragoza, Spain}

\affiliation{Instituto de Biocomputaci\'on y F\'{\i}sica de Sistemas
Complejos, Universidad de Zaragoza. 50018 Zaragoza, Spain}

\date{\today}

\begin{abstract}
This paper studies the features of a homopolymer translocating through a
flexible pore. The channel is modeled as a monolayer tube composed by monomers with two elastic parameters: spring-like two body interaction and bending three body recall interaction.  In order to guarantee the stability of the system, the membrane is compounded by a lipid bilayer structure having hydrophobic body (internal), while the pore is hydrophilic in both edges.
The polymer is end-pulled from the \emph{cis}-side to the \emph{trans}-side by a cantilever, to which is connected through a spring able to measure the force acting on the polymer during the translocation. All the structure reacts to the impacts of the monomers of the polymer with vibrations generated by the movement of its constituent bodies.
In these conditions, the work done by the cantilever shows a nonmonotonic behavior with the elastic constant, revealing a resonant-like behavior in a parameter region.
Moreover, the force spectroscopy registered as a function of time, is able to record the main kinetics of the polymer progression inside the pore. 

\end{abstract}

\pacs{87.15.-v, 36.20.-r, 87.18.Tt, 83.10.Rs, 05.40.-a}
\keywords{Stochastic Modeling, Fluctuation phenomena, Polymer dynamics, Langevin equation}


\maketitle

\section{Introduction}
Polymer translocation consists in the passage of an extended molecule through either artificial or biological channels. 
It concerns many biological processes as well as, nowadays, artificial manipulations at the nanoscale~\cite{KasPNAS96,NL,Merchant}.

In biology, DNA, RNA and protein translocates through cell membranes and nuclear pores,  the DNA is injected by the phages viruses, drugs are delivered to the cells through translocation processes, and also DNA sequencing can be performed with the help of the passage of the chain of nucleic acids through biological or solid state pores~\cite{RMP}. 

A variety of different models have been introduced to describe and study this and related problems
\cite{Meller2003,Milchev2011}, with some unified visions also proposed~\cite{Grosberg2, Ikonen-scaling-PRE2012}.
In simple models, a single barrier potential, also depending on time, is considered~\cite{Pizz2010,Pizz2013,Sung1996}; In others, attention has been devoted to stochastic ratchet-like forces~\cite{linke,linke2} to drive the process. 

The transport phenomena involves not only translocation through passive channels (as in artificial pores), but also active ones, eventually extended, that are modeled by introducing periodic sinusoidal forces~\cite{ajf-sin,golest2011,ikonen2012,golest2012}, either stochastic random telegraph noises (RTN)~\cite{ajf-rtn,ikonen2012}
or dichotomous ATP-based motor noises~\cite{ajf-damn,pffs-damn,afj-3d,afj-SR}, the latter motivated by the action of molecular motors as in the DNA injection/ejection in some types of viruses~\cite{Bust,Bust09}. 

In other examples the active pore effects have been  modelled through end-pulled mechanisms, provided experimentally by new devices as optical and magnetic tweezers \cite{Sara2018,Sara2014,Sara2015,Sara2017,Sara2017-SR,af-fs}.

The above-mentioned models are mostly based on modifications of the Rouse chain~\cite{Rouse} in the 1d to 3d domains.  Generally, they consider rigid and/or structureless pores, and usually they deal with externally provided driving forces, also in the case of a specific time dependence allowed inside the pore. Thus, poor attention has been reserved to the spatial dependence of the driving along the pore. 
In this sense, the understanding of the polymer translocation is still incomplete, and various aspects of the real translocation phenomenon deserve deeper attention. 
For example, recent works suggest that the translocation process significantly depends on the size and flexibility of the
chain~\cite{slater2013,larrea2013}, and on interactions between the polymers and the pore~\cite{Menais2016}.

\begin{figure}[tb]
  \centering
  \includegraphics[width=8.5cm]{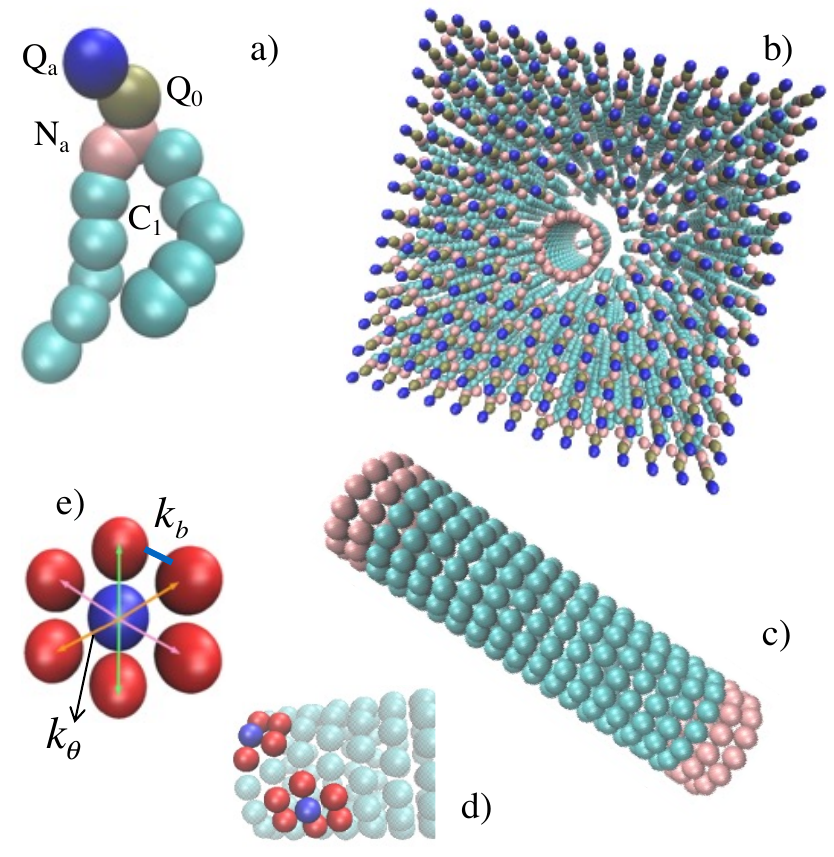}
  \caption{Model of the membrane as compound by a bilayer structure of lipid molecules (DSPC), and the channel formed one cylindrical-shaped layer of monomers with two rings of hydrophilic monomers and both two-body and three-body elastic interactions.} 
  \label{schema}
\end{figure}

Aim of this work is to investigate the spatio-temporal pore-membrane reaction during the translocation as a function of the elastic properties of the pore, by using a semi-realistic approach.
Some heuristic attempt in introducing the spatio-temporal pore reaction has been recently done in refs.~\cite{Ceconni,Rubi}, and with a phenomenological effective potential in ref.~\cite{shulten}.

Here, we consider the case of a polymeric molecule in the 3d domain
driven by pulling it at a constant velocity through a cantilever attached to one edge by means of a spring which permit to measure the force being applied during the process. Though maintaining a mesoscopic description, not only the polymer but also both the channel and the membrane are modeled as a many body compound which move in a thermal bath.
Consequently, the pore and the membrane fully participate to the translocation dynamics as well as the polymer itself. 

We will specially focus on the study of the force registered in the spring, and analyze the basic movements of the pore by means of the principal component analysis (PCA) of the trajectories, which is able to reveal, as a function of the flexibility parameters, the most important kinetic modes of the pore involved in the translocation.

Our study is not only relevant by its basic interest which allows to enlarge the modelistic view of the translocation, but also because of its relation to biological systems. One nice example is again the ejection of DNA virus through the pores of the capsid, where it has been seen that the mechanical properties of the capsid modify the pore elasticity to favor the DNA translocation~\cite{Mateu}. 

The article is organized as follows: first we present the model of the polymer, of the membrane and the pore inside it.  In Sec.~III we will show the main features of the force registered at the cantilever and the revealed translocation steps of the polymer inside the pore. In Sec.~IV, before the conclusion section,  we analyze the vibrational modes of the pore both without the polymer as well as during the translocation.

\section{The model}

The model consists in three different structures: the membrane, the channel, and the polymer. The molecule is represented as an homomeric chain formed by $N$ monomers. The monomer's interactions includes Hooke's longitudinal elasticity, bending energy, and excluded volume effects and interaction with the monomers, the membrane and the pore. 

The elastic potential energy is given by
 \be V_{\rm el}(d_i)=\frac{k_b}{2}\sum_{i=1}^{N} (d_i-l_0)^2,
 \label{v-har}
 \ee
\noindent where $k_b$ is the elastic parameter, $\mathbf{r}_i$ is
the position of the $i$-th particle, $d_i = |\mathbf{d}_i| =
|\mathbf{r}_{i+1}-\mathbf{r}_i|$ is the distance between the
monomers $i$ and $i+1$, and $l_0$ is the equilibrium distance between adjacent
monomers.

The bending energy of the chain with a term given by
 \be
  V_{\rm ben}(\theta_i)=\frac{k_{\theta}}{2}\sum_{i=1}^{N} [1-\cos(\theta_i-\theta_0)],
 \label{v-ben}
 \ee
where $k_{\theta}$ is the three bodies bending elastic constant, $\theta_i$ is the
angle between the link $\mathbf{d}_{i+1}$ and the link $\mathbf{d}_{i}$, and $\theta_0$ the equilibrium angle, $\theta_0=180^{\circ}$ in our case.

The excluded volume is modeled as a repulsive only Lennard-Jones potential (sometimes called Weeks-Chandler-Andersen potential):
\be V_{\rm LJ}(r)
  = 4\epsilon \left[ \left(\frac{\sigma}{r}\right)^{12}-\left(\frac{\sigma}{r}\right)^6 \right]+\epsilon
 \label{LJ}
 \ee
for $r\leq 2^{1/6}\sigma$, and zero otherwise.

The dynamics of every monomer of the chain is executed by using the GROMACS environment~\cite{Gromacs}, and the described potentials are part of the MARTINI force fields~\cite{Martini,Martini2}, which has permitted the above-descripted mesoscopic reduction. 

The parametrization of the polymer partially follows the Polyethylene glycol (PEG) features in the Martini force fields, with the care to avoid the attractive Lennard-Jones hydrophobic interactions between the monomers, by using the potential of Eq.~(\ref{LJ}), manually provided in the force field context~\cite{Martini}. This way model an extended polymer instead of a compact one, whose study is left to a future investigation.

The membrane has been built by means $N_{Phos}=368$ units of phospholipids ({\it Distearoylphosphatidylcholine}, DSPC) modeled by means of 4 different types of monomers (See Fig.~\ref{schema}a). The head of the phospholipids (blue monomer) present a hydrophilic interaction, while the legs, compounded of 5 monomers each, has a hydrophobic interaction with a string attraction between each other, so guaranteeing the formation of the bilayer membrane structure (See Fig.~\ref{schema}b)). 

The pore is built by means of a connected sequence of monomers which interacts each other by means of the three potentials just described, and constitute, as the polymer itself, a single molecule. The edges of the channel are compounded by two rings of hydrophilic monomers, while the inner body is compounded by hydrophobic particles. This way the channel can remain stably inserted into the membrane structure.

In this context, the simulation provided consist in the numerical solution of a classical Langevin equation:

 \bea
 m\ddot{\mathbf{r}}_i + m \Gamma\dot{\mathbf{r}}_i &= &- \mathbf{\nabla}_i V_{\rm el}(d_i)
-\mathbf{\nabla}_i V_{\rm ben}(\theta_i) -\mathbf{\nabla}_i V_{\rm
LJ}(d_i) \nonumber \\ &+& F_{drv,i} \mathbf{i} + \mathbf{F}_{sp,i}
+ \sqrt{2m\Gamma k_BT} \vec{\xi}_i(t),
 \label{eq}
 \eea

where  $\Gamma$ and $m$ are the damping parameter and mass of each monomer respectively.  $\vec{\xi}_{i}(t)$ stands for the Gaussian uncorrelated thermal fluctuation and follows the usual statistical properties $\langle\xi_{i,\alpha}(t)\rangle=0$ and $\langle\xi_{i,\alpha}(t)\xi_{j,\beta}(t')\rangle = \delta_{i
j}\delta_{\alpha,\beta}\delta(t'-t)$, with $i=1,...,N$, and
$\alpha = x,y,z$. The operator $\mathbf{\nabla}_i =
\partial / \partial x_i \, \mathbf{i} + \partial / \partial y_i
\, \mathbf{j}  + \partial / \partial z_i \, \mathbf{k}$.

The two explicit forces in Eq.~(\ref{eq}) account for the driving forces
over the particles inside the pore $F_{drv}$, and the chain-membrane and chain-pore spatial constraints $\mathbf{F}_{sp}$. This latter is modeled by using the
same repulsive Lennard-Jones potential described in Eq.~(\ref{LJ}) and acts between any couple of monomers, being them part of the channel or of the membrane, or of the polymer itself. 

$F_{drv}$ is the Hooke's force acting on right-end edge of the polymer, being the cantilever pulled rightward while connected with a spring to the last monomer of the chain.
To complete this section we list the parameters used in the simulations.  The
temperature used is $T=300$K. The pore length is $l=6.6nm$ (that corresponds to $N \approx20$ monomers), the rest distance between the monomers is $l_0=0.33nm$, which is the same for the pore and the polymer, the polymer length is $l_{pol}=30nm$ long (that corresponds to $N=90$ monomers) and the Lennard-Jones parameters are $\epsilon=5.6kJ/mol$, and $\sigma=0.47nm$.

The elastic values are different for the different parts of the system. For the polymer, they remain fixed along the work: $k_{b,pol} = 1.7 \cdot 10^4 kJ/nm^2/mol$,  $k_{\theta,pol} =10 kJ/mol$, with equilibrium angle $\theta_{0} = 180^{\circ}$, while the pore values are: $k_{b} = 7500 kJ/nm^2/mol$,  $k_{\theta} =50 kJ/mol$. These two latter values, which we write here as a reference, are the Martini constants for the {\it alanine amino acid}, but they will varied in the subsequent study. Finally the parameters for the DSPC are the standard ones in the Martini environment \cite{Martini}.

\section{Polymer translocation}
 \begin{figure}[b]
 \centering
 \includegraphics[angle=0, width=8.5cm]{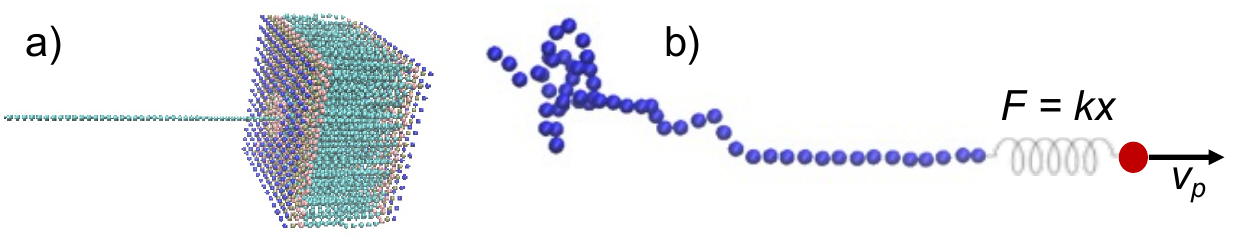}
 \caption{a) Initial configuration before the thermalization. The head of the polymer starts from the middle of the pore. b) Pull scheme with the cantilever and the spring connected to the first monomer.}
 \label{INI}
 \end{figure}
The main goal of this work, is to verify the effect of the pore elasticity on the translocation.
We performed a set of numerical experiments by spanning in both the bonding elastic parameter $k_b$ and the bending $k_{\theta}$ by setting the pulling velocity to $v_{p} = 0.1\,nm/ps$, and the damping to the low value $\Gamma = 0.1 ps^{-1}$.  
 \begin{figure}[tbp]
 \centering
 \includegraphics[angle=0, width=0.48\textwidth]{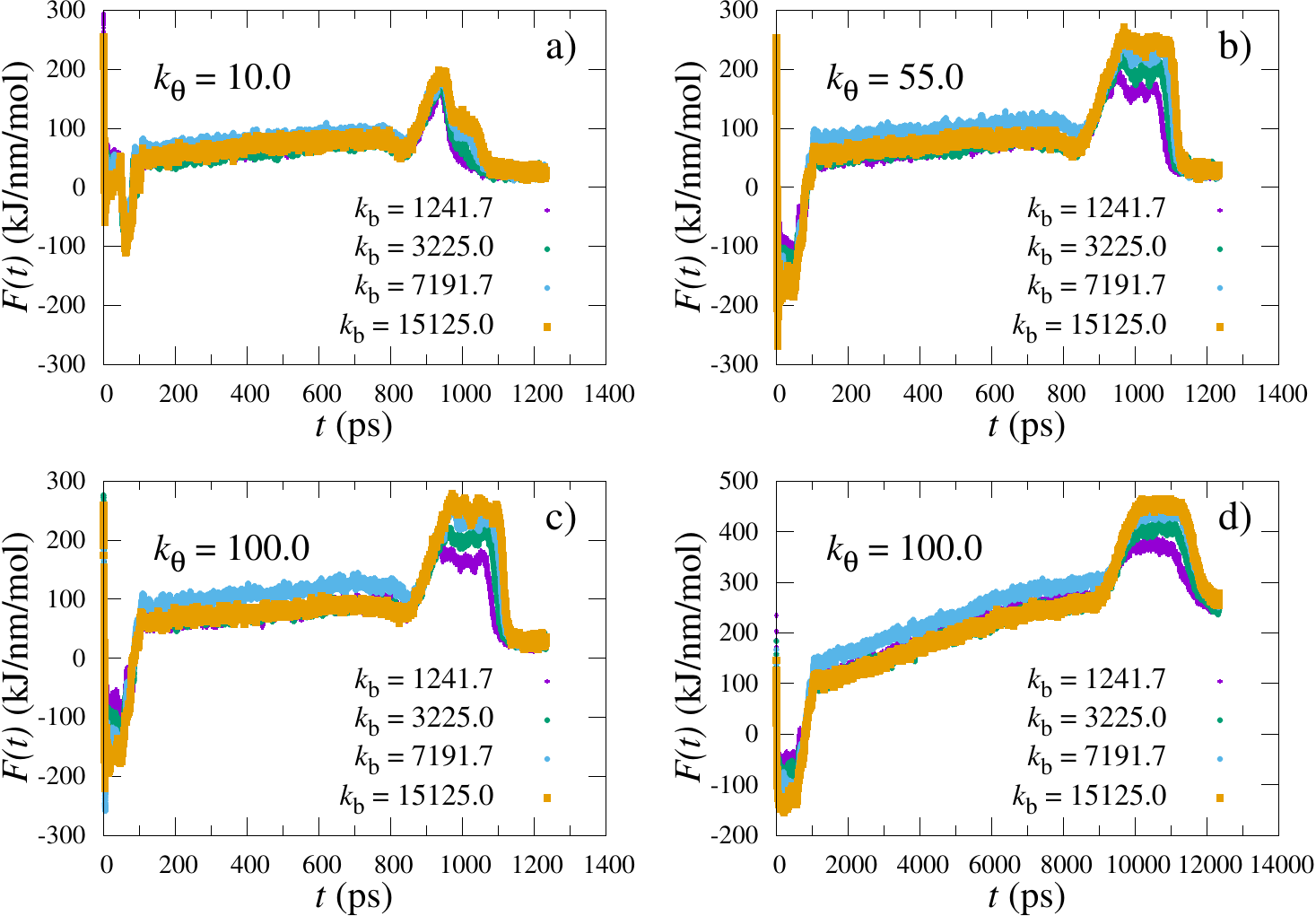}
 \caption{Average trajectories of the force registered at the cantilever as a function of time for the polymer translocation pulled at $v_{p}=0.1\,nm/ps$, with $\Gamma = 0.1 ps^{-1}$, at $T=300$K.  The last panel d) shows an example of trajectories at $k_{\theta}=100\,kJ/mol$, with $v_{p}=0.01\,nm/ps$ and $\Gamma =10 ps^{-1}$, whose results are qualitatively very similar to panel c), with an higher force at the cantilever due to the higher damping $\Gamma$ and a longer translocation time due to the reduced pulling velocity $v_p$.}
 \label{Trajectories}
 \end{figure}

The initial configuration in the translocation consists in the right edge of the polyer put in the middle of the pore channel.  Starting from that position (see Fig.~\ref{INI}a)), and maintaining fixed the position of the cantilever, the dynamics is run during a fixed time under the action of the thermal fluctuations, with the goal to reach a thermalization of the system, involving all the structures, i.e. polymer, membrane and channel. We have also estimated the relaxation time required for the polymer to reach the equilibrium state to be about  $\tau_{relax}\approx1\,\rm{ns}$. This time is of the order of the translocation time for one of the set of parameters used (See Fig.~\ref{Trajectories}a)b)c)). For other set of parameters, $\tau_{relax}$ is much smaller than the translocation time (See Fig.~\ref{Trajectories}d)), so showing a quasi-equilibrium kinetics. In all cases the curves remain qualitatively unchanged for the purpose of characterizing the position of the polymer during the translocation and its effects on the dynamics of the pore.

After that transient, the cantilever is moved rightward with the constant velocity $v_{p}$ (See the scheme in Fig.~\ref{INI}b)), in agreement with many experimental procedures, and the force on the connected spring is recorded up to the final event, defined as the last monomer of the chain reaching the right edge of the pore.

Some forces recorded at the spring as a function of the time for different elastic parameters are reported in Fig.~\ref{Trajectories}, averaged over 48 realizations. We use the parameters $v_{p}=0.1\,nm/ps$ and $\Gamma = 0.1 ps^{-1}$, at $T=300$K as mainly imployed in the next analysis, with one example, for comparison, for $k_{\theta}=100kJ/mol$ with $v_{p}=0.01\,nm/ps$ and $\Gamma = 10 ps^{-1}$. By using this latter set of parameters, the qualitative behavior remains analogue as the former set also in the cases not here shown. In all cases, it can be observed that the force increases for large part of the polymer translocation up to an abrupt peak, that is followed by a final reduction of the force recorded.  The features evidenced in the force evolution barely depend on the bond elastic parameter $k_b$, which just provides a slightly higher force as $k_b$ increases, but they do depend on the bending constant $k_{\theta}$, showing a saturating behavior with increasing $k_{\theta}$ with an increase of the force intensity, and an enlargement of the force peak at the end of the dynamics. 

\begin{figure*}[tbp]
  \centering
  \includegraphics[angle=0, width=0.8\textwidth]{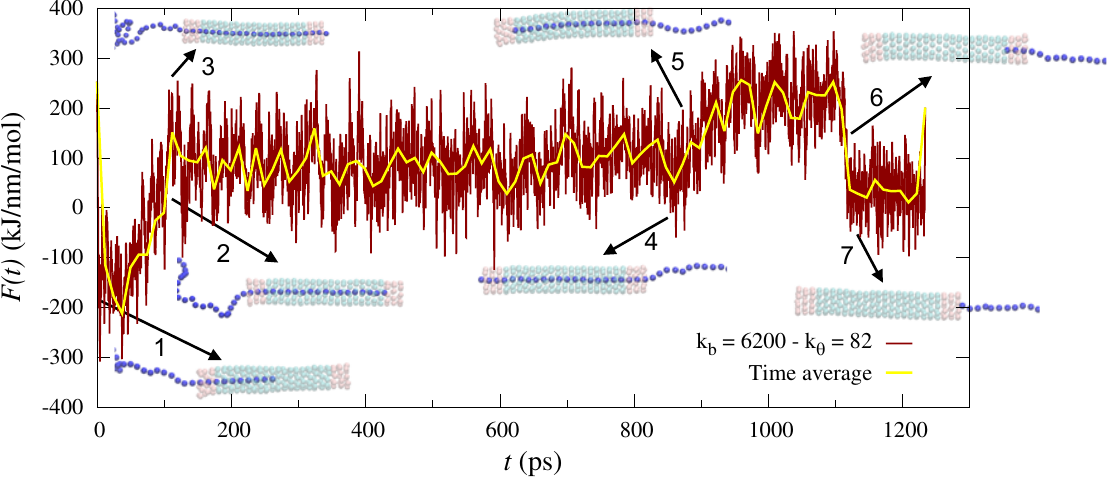}
  \caption{Individual force trajectory as a function of time during the translocation. The seven thresholds show the position of the polymer in specific points of the channel as commented in the text.} 
  \label{TrajectoryDetails}
\end{figure*}
\subsubsection{Trajectory details}

The forces as a function of the time shown in Fig.~\ref{Trajectories} follow in all cases a clear pattern.
Fig.~\ref{TrajectoryDetails} shows one example of these force trajectory which is able to detect different regions reached by the polymer during the translocation. The features shown in the force can be understood remembering that the pore and the polymer are compounded by not neutral beads. On the contrary, we distinguish different monomers in the pore: the edges of the pore, which corresponds to the external sides of the membrane, are hydrophilic, while the inner monomers of the pore are hydrophobic. This way the pore can maintain inside the membrane without additional bonding degrees.
According to this variety, the monomers of the translocating chain, interact differently with the different monomers of the pore channel, and that explains the structured shape of the $F(t)$ trajectories, that are characterized by different thresholds which correspond to the times the polymer reaches specific positions inside the pore. Fig.~\ref{TrajectoryDetails} presents those thresholds as arrows with a numbered label whose meaning is commented in the following list. 
The first part of the trajectories presents a negative outcome of the force. This means that the polymer, which starts with its head at the middle of the pore, feels a \emph{rightward attractive effective force}. This effect is due to the specific interaction between the pore and the polymer. In fact the interaction between the monomers of the chain and the pore is hydrophobic, \emph{i.e.} it can be slightly attractive as well as the body of the pore. Being the polymer not symmetrically distributed inside the pore at the beginning, the right part of the pore attracts the polymer to the right, up to the condition in which the polymer is symmetrically distributed inside. As a result the force recorded at the cantilever is initially strongly negative. This condition occurs around the threshold $1$ here below indicated, where all the threshold drawn in Fig.~\ref{TrajectoryDetails} are described: 
\begin{enumerate}
\item The kinetics starts with the head of the polymer in the middle of the pore and the force becomes rapidly negative.
\item The head of the polymer approaches the right hydrophilic edge of the pore. 
\item The head of the polymer reaches the \emph{trans} region (the polymer is now symmetrically distributed inside the pore). 
\item The polymer tail reaches the pore entrance. 
\item The polymer tail overcomes the left hydrophilic pore edge and reaches the hydrophobic region inside the pore. The polymer distribution is now asymmetrical, opposite with respect what happened at the beginning, and the polymer is pulled {\it leftward} by the pore hydrophobic body, so increasing the forces registered at the cantilever. 
\item The polymer tail reaches the right edge. 
\item The polymer tail exits the pore. 
\end{enumerate}
After the tail exit, the trajectory only register a drag force no more influenced by the movements of the pore.

\subsubsection{Work span and Mean translocation time}
The repetition of $N_{exp}=100$ simulations with the same initial starting state, allows the measure of the mean translocation time, i.e. the ensemble average $\tau$ of the time $t_i$ spent by the polymer to totally enter in the \emph{trans} region of the membrane in each experiment,  and the average work done by the cantilever, $W = W_i/N_{exp}$ with $W_i = \int F_i(x) dx$, $F_i(x)$ the force registered at the cantilever, and $i$ indicates the specific realization with $i=1...N_{exp}$.

\begin{figure}[tbp]
 \centering
 \includegraphics[angle=0, width=0.48\textwidth]{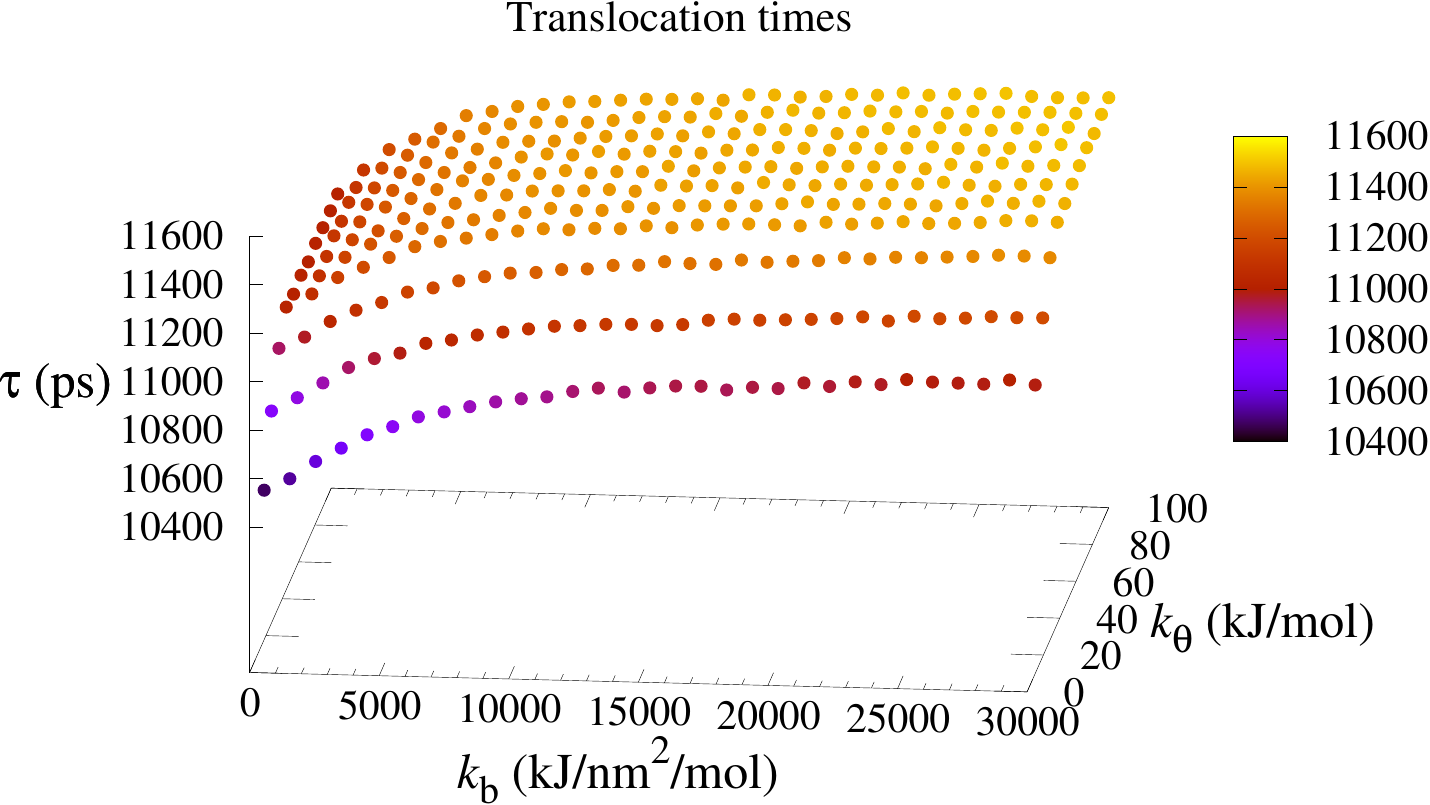}
 \caption{Mean translocation times as a function of the elastic parameters  $k_b$ and $k_{\theta}$}
 \label{Map_TT}
 \end{figure}

The mean translocation time reported in Fig.~\ref{Map_TT} show a smooth increase by increasing both the elastic parameters, $k_b$ and $k_{\theta}$. In this simulations, the mean translocation time appears a monotonic magnitude, being the translocation time mainly imposed by the cantilever which moves rightward at a fixed velocity $v_p$. In general, more flexible membranes, in both the bonding and bending parameters, give rise to lower translocation times.  

\begin{figure}[tbp]
 \centering
 \includegraphics[angle=0, width=0.48\textwidth]{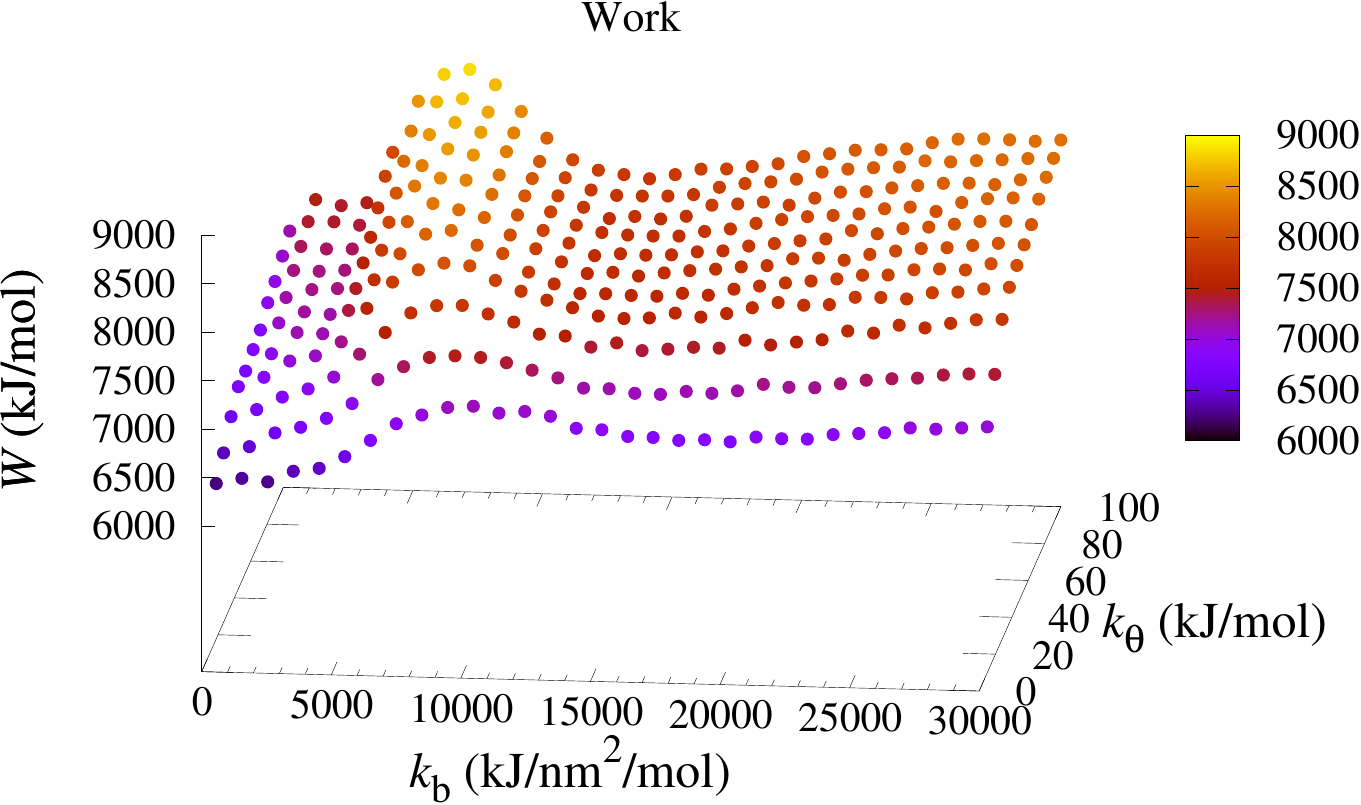}
 \caption{Mean work done by the cantilever in the translocation as a function of the elastic parameters  $k_b$ and $k_{\theta}$.}
 \label{Map_work}
 \end{figure}

More interesting appears the average work presented in Fig.~\ref{Map_work}, especially if we focus on the $k_b$ parameter. In fact, the work increases monotonically as the bending parameter $k_{\theta}$ increases, while at any fixed $k_{\theta}$ values, a clear maximum of the work is visible as a function of the parameter $k_b$. Both the results, work, and translocation time as a function of the elastic coefficients, have been shown here by using the parameters $v_{p}=0.01\,nm/ps$ and $\Gamma =10 ps^{-1}$, and the qualitative behavior is identical also when using the parameters $v_{p}=0.1 \,nm/ps$ and $\Gamma =0.1 ps^{-1}$.

This maximum is due to the resonance of the system with the inner degree of freedom of the structure as we will show in the next section.

\section{Dynamical features of the pore}

The results of the translocation put in evidence the nonmonotonic behavior of the work as a function of the elastic properties of the pore, as well as the enlarged maximum of the time trajectory of the forces (See Fig.~\ref{Trajectories}), rise the hypothesis that some kind of resonances appears in the interaction pore-polymer during the translocation.
 In order to find such kind of resonances, we study the effects of elasticity features on the dynamics, by performing the Principal Component Analysis (PCA) using the deterministic and stochastic dynamics, which are equivalent to zero and room temperature, respectively. We will consider both the cases: with and without polymer translocation.

\subsection{Principal Component Analysis - pore alone}

The PCA is a methods able to enormously reduce the complexity of a trajectory in order to give informations only about the displacement of the main degree of freedom of the structure~\cite{PCA}.
The method takes into account the covariance matrix of the time average of the trajectory coordinates having rested the corresponding CM at every time step, so accounting for the main inner displacement.
The covariance matrix is defined as 
\be 
 C_{i,j} =  \langle \zeta_i - \langle \zeta_i \rangle \rangle_t  \langle \zeta_j - \langle \zeta_j \rangle \rangle_t 
 \label{Cov}
 \ee
where $\langle \cdot \rangle_t$ indicates the average over all the times of the selected coordinates, and $\zeta = (x,y,z)$, and so $i,j=1,...,3N$, with $N$ is the number of particles of the pore. The diagonalization of this matrix gives rise to a 3N eigenvalues $\lambda_i$, that can be order from the bigger to the smaller, corresponding to an increasing frequency of the oscillations. The movement can then be projected to the eigenvectors $\vec{v}_i$ which correspond to ``normal" movements which are somehow equivalent to the normal modes of oscillations. Thus, the $i$-th principal component (PC) trajectory is defined as
 \be 
 PC_{i}(t) =  \vec{\zeta}(t)  \cdot \vec{v}_i.
 \label{pc}
 \ee
Differently than the normal modes, PC trajectories are a straightforward approximation of the real motion, and allow a computational analysis independently on the knowledge of the analytical potential functions involved in the dynamics. Moreover, PCs reduce to the normal modes at low temperatures. 
\begin{figure}[b]
  \centering
  \includegraphics[width=0.48\textwidth]{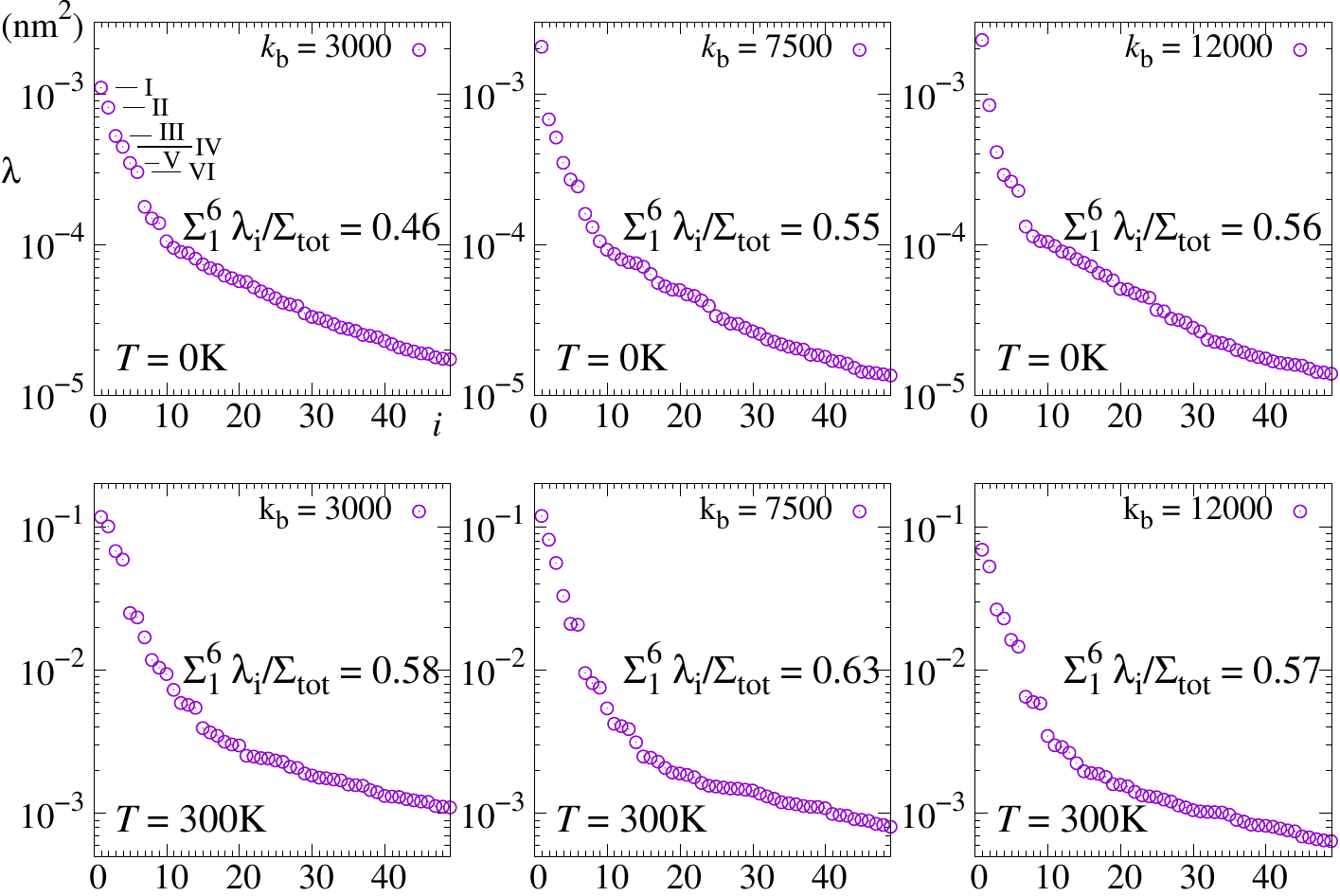}
  \caption{PCA eigenvalue distribution for $k_b=3,000, 7,500$  and $12,000 \, kJ/nm^2/mol$ with both the \emph{pore alone} and $T=0$K (1$^{\rm st}$ row) and \emph{during the polymer translocation} at temperature with $T=300$K (2$^{\rm nd}$ row). The first 6 eigenvalues of the first panel define de PCA eigentrajectories with labels I--VI whose kinetics are reported in Fig.~\ref{modes}.} 
  \label{PCAEigen}
\end{figure}
The PCs have been computed for trajectories with different values of the elastic parameter of the pore $k_b$ and $k_{\theta}$, with both temperatures: $T=300$K and $T=0$K. In this latter simulations, the system has been thermalized at $T=300$K, and then left to relax at $T=0$K in order to show the presence of preferred movements of the channel during the transient.  This way the dynamics detected is the one related to the inner displacements of the structure, even without stochastic forcing. In other words, the PCA takes into account the transient movement before relaxing, and identifies in this time the most relevant oscillatory components.
Fig.~\ref{PCAEigen} presents the plots, in the descendent order, of the eigenvalues $\lambda$s obtained in the analysis, for the pore alone at $T=0$K (1$^{\rm st}$ row), and during the translocation at $T=300$K (2$^{\rm nd}$ row). It can be seen as the first 6 are generally the most important ones because the $7^{\rm th}$ generally appears after a gap from the $6^{\rm th}$. This group is then the most representative contribution to of the complete movement, and the sum of the six firsts eigenvalues correspond to a minimum of 45$\%$ of the total contribution, as shown inside the plots in the figure.
\begin{figure}[t]
  \centering
  \includegraphics[width=0.95\columnwidth]{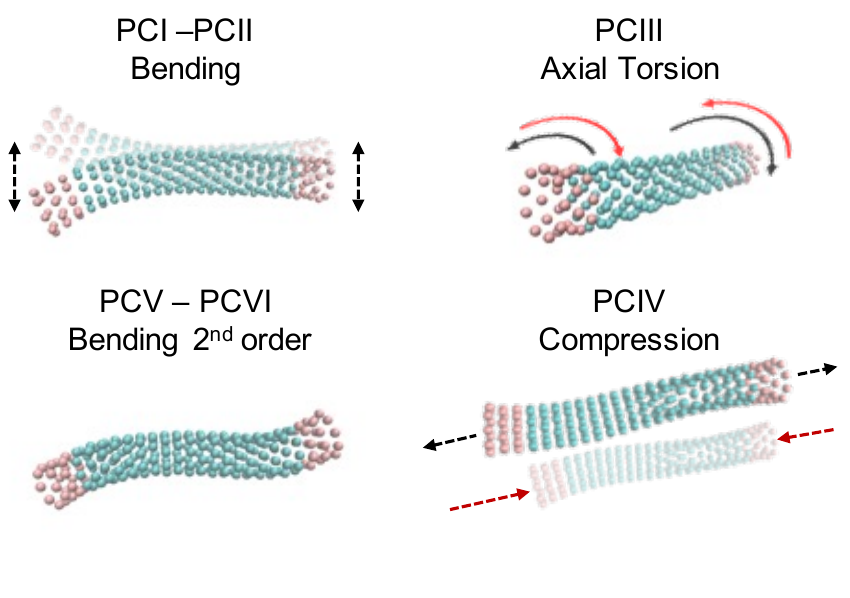}
  \caption{Examples of the most important movements of the pore according to the PCA.  The latin numbers represent the position of the mode in the decrescent eigenvalue's list in the zero temperature case and small $k_b$ values. The two PCs I and II refer to the bending movement in the $(x,z)$ and $(y,z)$ planes respectively.  The same also applies to the two 2$^{\rm nd}$ order bending modes PCV and PCVI.} 
  \label{modes}
\end{figure}

Fig.~\ref{modes} shows the most relevant PCs movements associated to each eigenvalue obtained by projecting the real trajectory to the corresponding eigenvector. We can observe the presence of the bending oscillations in the two directions orthogonal to the $z$-axis (and orthogonal between them), which we call PCI and PCII, the torque oscillation in the $z$-direction (PCIII), the longitudinal compression/expansion along the $z$-direction (PCIV), and the bending 2$^{\rm nd}$ harmonic oscillations (PCV and PCVI).
 
\begin{figure*}[tb]
  \centering
  \includegraphics[angle=-0, width=0.95\textwidth]{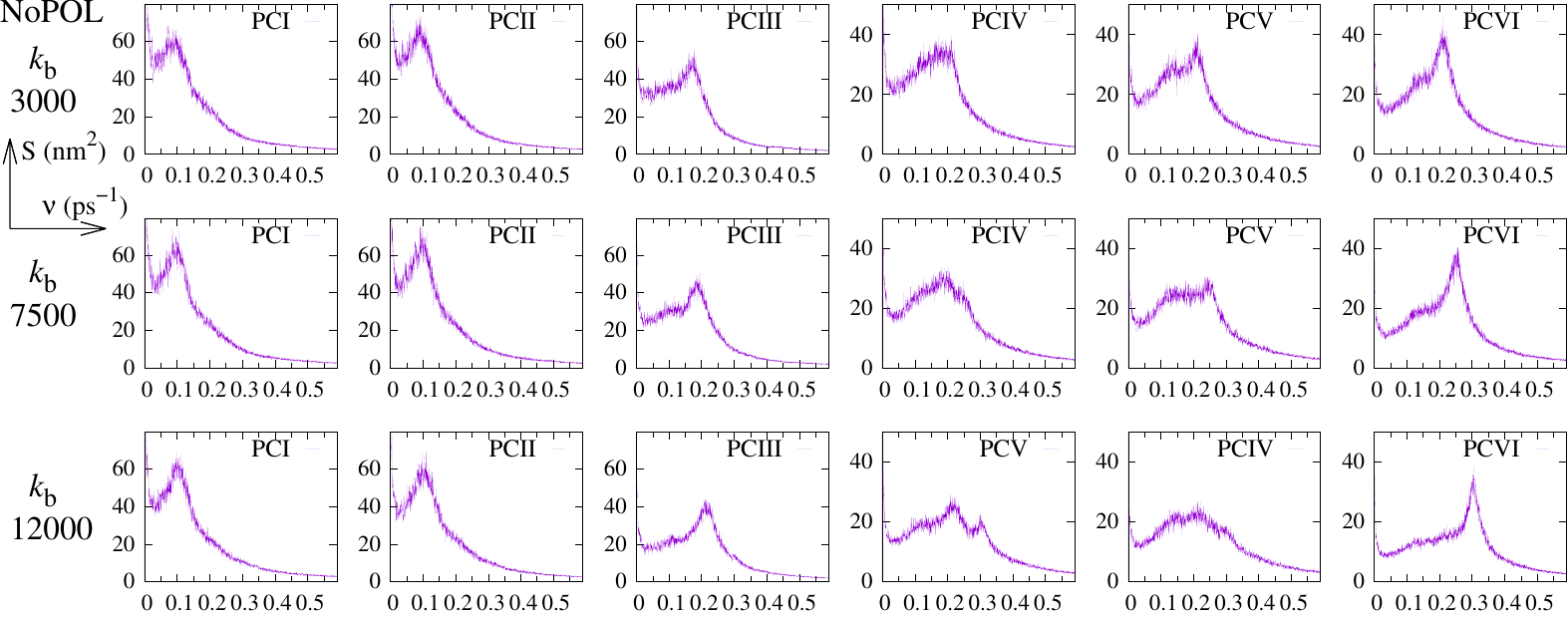}
  \caption{The Fourier transform (power spectrum) of the six most relevant eigentrajectories of the PCA (PCI-PCVI), with the pore alone, for $k_b = 3,000, 7,500 \,\,\rm{ and } \,\,12,000 \, kJ/nm/mol $, i.e. the elastic values corresponding to the work before, at, and after the resonant-like behavior detected in the translocation process and shown in Fig.~\ref{Map_work}. In all simulations $k_{\theta}=50 kJ/mol$.} 
  \label{PCA_T0SP}
\end{figure*}

Fig.~\ref{PCA_T0SP} shows the Fourier transform (power spectrum $S(\nu)$) of the principal components for $k_b = 3,000, 7,500$,  and $12,000 \,kJ/nm^2/mol$, \emph{i.e.} the elastic values corresponding, respectively, to the points located before, at, and after the resonant behavior of the work represented in Fig.~\ref{Map_work}. We can notice how the frequency of the longest vibrational modes  are almost unmodified by varying the elastic parameter $k_b$. In fact, the PCI and PCII Fourier transform do not change by changing the three parameters $k_b$. Conversely the vibrational modes PCIII (torque), PCIV, PCV (bending 2$^{\rm nd}$ order) and PCIV (compression), do change by increasing the $k_b$ parameter in both the displacement of the maximum, and the increase of the width of the corresponding Fourier transforms. These features are those one can expected when increasing  the rigidity of the pore.

\subsection{Principal Component Analysis - with polymer}

The subsequent analysis is oriented to investigate the changes in the dynamics of the pore $during$ the polymer translocation, with the goal to find which vibrations features of the kinetics of the pore alone survive and/or are excited, so justifying the resonant-like behavior of the work.

For this purpose we have performed an analysis of the polymer translocation at a relatively low damping ($\Gamma=0.1\,ps^{-1}$)  to evidence, through the Fourier analysis of the PC of the kinetics of the pore, the vibrational frequencies activated during the process.

\begin{figure*}[tb]
  \centering
  \includegraphics[width=0.99\textwidth]{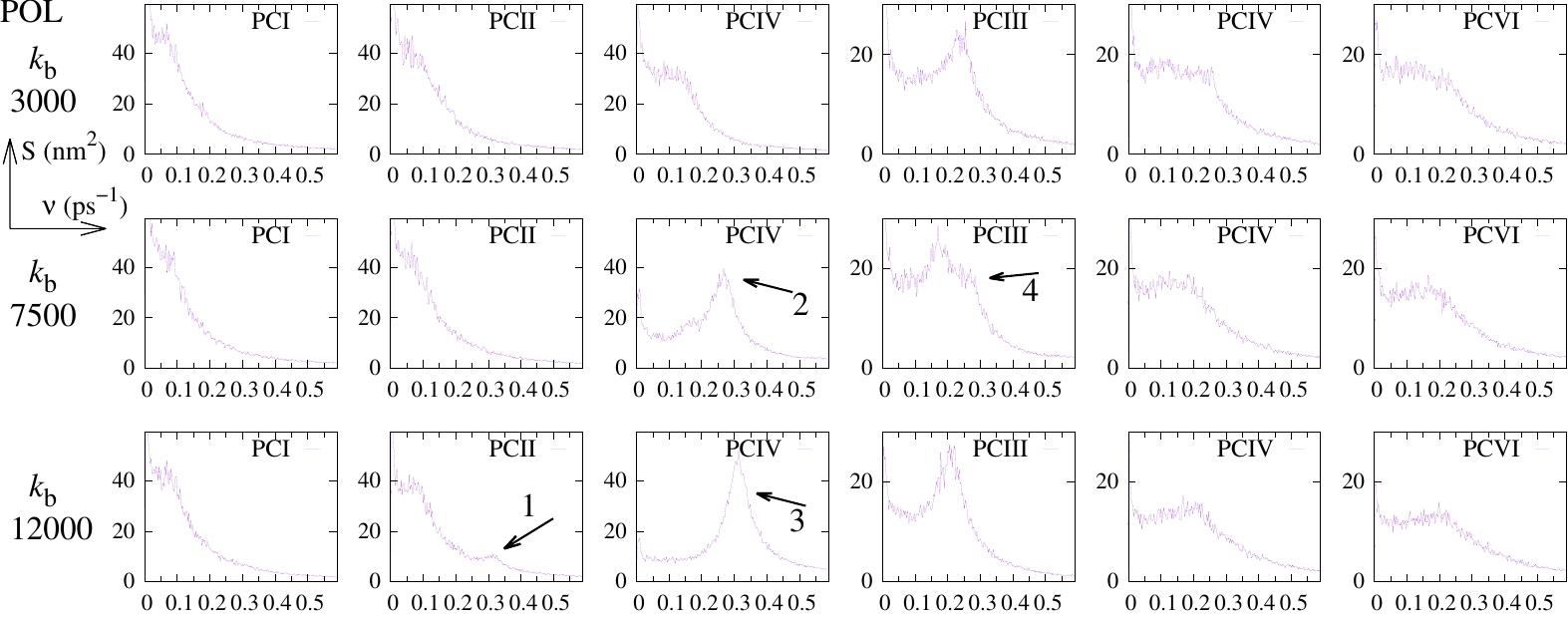}
  \caption{Fourier transforms (power spectrum) of the eigentrajectories corresponding to the principal components PCI-PCIV as in Fig.~\ref{PCA_T0SP} at the same elastic values of the bond constant $k_b=3,000, 7,500 \,\,\rm{ and } \,\,12,000 \,kJ/nm/mol$ \emph{during the polymer translocation}, at temperature $T=300$K, damping $\Gamma =0.1\,ps^{-1}$, $v_p =0.1\,nm/ps$, averaged over 48 simulations.} 
  \label{PCA_T300P}
\end{figure*}

The results of these simulations are plotted in Fig.~\ref{PCA_T300P} again for the three longitudinal elastic constant $k_b=$ 3,000, 7,500 and $12,000  \,kJ/nm^2/mol$ above used, where the Fourier transform for the most important (highest eigenvalues) vibrational movement of the principal components previously selected by the pore PC analysis above shown. Given the high stochastic character of these simulations, the trajectories have been averaged over $N_s=48$ realizations in order to take into account the presence of thermal fluctuation, as the temperature of the translocation has been set to $T=300$K. 

We can observe that the pulling induces the activation of the compression vibrational mode. In fact, with respect to the pore analysis ($T=0$K), where the third most important vibrational mode is the torsional kinetics (PCIII), in the presence of the pulled polymer this mode just remains predominant for $k_b=3,000$, but for $k_b=7,500$ and $k_b=12,000 \,kJ/nm^2/mol$, the compression mode PCIV becomes more relevant than PCIII (See the panels with the arrows 2 and 3 in Fig.~\ref{PCA_T300P}). In addition, the peak frequency of this compression mode increases to the value $\nu\approx 0.33 \,ps^{-1}$, that corresponds to the effective forcing frequency excerced by the monomers in their interaction with the pore entrance during the translocation. In fact, given the distance of the monomers in the pore be $l_0$ and considering the translocation velocity approximately equal to the pulling velocity $v_p$, the frequency of the monomer entrance in the pore results is given by $\nu_t \approx v_p/l_0 \approx 0.33\, ps^{-1}$. The compression/elongation mode of the channel is then synchronized with the pore vibrations in the case of rigid bond interaction (high $k_b$ values), and results in an increase of the work necessary for the translocation. 
Moreover, the same vibrational mode is also weakly mixed in the $2^{\rm nd}$ PC for $k_b=12,000\, kJ/nm^2/mol$, where a small peak is present at around $\nu\approx 0.33\,ps^{-1}$ (arrow 1 in the figure). 

The polymer pulling also affect another eigenfunctions of the movements. The fourth column of Fig.~\ref{PCA_T300P} corresponds to the torsional vibration modes which results perturbed by the presence of, again, of the compression mode which weakly appears as frequency bump in the fourth panel of the elasticity $k_b=7,500\,kJ/nm^2/mol$ (see the arrow 4 in the figure).

Summarizing, we interpret all these changes as the effects of the polymer translocation which induces a periodic forcing according to the monomers entering the pore. Their interactions with the pore entrance, with an approximately constant frequency, induces a compression/enlargement movement that becomes more relevant than the torsional mode as obtained from the order of the eigenvalues, and, in addition, mixes the bending and torsional modes with the compression one. Specifically, the torsional mode synchronizes with this movement for a specific elasticity region, so requiring the higher energy administered to the pore,  which generates the resonant-like phenomenon of the work described above. 

\section{Summary and Conclusions}
This work describes the translocation features of an end-pulled polymer through a many body pore inserted into a membrane by using a mesoscopic approximation with the minimal potentials present in real polymer chains.

The hydrophilic and hydrophobic features, together with the elastic properties of a real system have been taken into account in this model in both the pore and the membrane. This choice allowed to explore the structured force trajectory during the translocation as well as the the mean values of the translocation time and the total work done to translocate. This latter has been found to show a resonant-like behavior with a strong maximum as a function of the longitudinal elastic parameter of the pore.

The resonant-like behavior has been put in relationship with the elasticity strength of the pore, explored through a principal component analysis, that showed the pore vibrating, with respect to the empty pore kinetics, in a different way in the presence of the polymer. Specifically, the translocation of the polymer excites the torsion and compression vibrational modes more for rigid structures than for soft ones.

The presented analysis shows that the internal degrees of freedom of the pore and the membrane do affects in a sensible way the kinetics features of the translocation. Beside an increase of the translocation time for more rigid structures, both in the bonding and in the bending parameters, the presence of resonant-like response in the work warns about the use of static structureless pores in the study of this process. 

This work represents the first contribution to a many-body -yet mesoscopic- description of the polymer translocation features, and gives a step forward toward the microscopic approach of long molecules translocation. 

\section*{Acknowledgments}
This work acknowledges the Grant PID2020-113582GB-I00 funded by MCIN/AEI/ 10.13039/501100011033.  We also thank the support of the Aragon Government to the Recognized group `E36\_20R F\'isica Estad\'istica y no-lineal (FENOL)'.

\end{document}